\documentclass[sigconf]{acmart}
\usepackage{tabularx}
\usepackage{booktabs}
\usepackage{multirow}
\usepackage{url}

\AtBeginDocument{%
  }

\setcopyright{acmlicensed}
\copyrightyear{2025}
\acmYear{2025}
\acmDOI{XXXXXXX.XXXXXXX}
\acmConference[SBES '25]{Brazilian Symposium on Software Engineering}{September 22--26, 2025}{Recife, PE}
\setcopyright{none}
\renewcommand\footnotetextcopyrightpermission[1]{}

\begin{document}

\title{Educational Insights from Code: Mapping Learning Challenges in Object-Oriented Programming through Code-Based Evidence}



\author{André Menolli}
\affiliation{
  \institution{Universidade Estadual do Norte do Paraná}
   \institution{Universidade Estadual de Londrina}
  \country{Brazil}
}
\email{menolli@uenp.br}

\author{Bruno Strik}
\affiliation{
  \institution{Instituto Federal do Paraná}
 \institution{Universidade Estadual de Londrina}
  \country{Brazil}
 }
\email{bruno.strik@ifpr.edu.br}

\renewcommand{\shortauthors}{Menolli and Strik}

\begin{abstract}
Object-Oriented programming is frequently challenging for undergraduate Computer Science students, particularly in understanding abstract concepts such as encapsulation, inheritance, and polymorphism. Although the literature outlines various methods to identify potential design and coding issues in object-oriented programming through source code analysis, such as code smells and SOLID principles, few studies explore how these code-level issues relate to learning difficulties in Object-Oriented Programming. In this study, we explore the relationship of the code issue indicators with common challenges encountered during the learning of object-oriented programming.  Using qualitative analysis, we identified the main categories of learning difficulties and, through a literature review, established connections between these difficulties, code smells, and violations of the SOLID principles. As a result, we developed a conceptual map that links code-related issues to specific learning challenges in Object-Oriented Programming. The model was then evaluated by an expert who applied it in the analysis of the student code to assess its relevance and applicability in educational contexts.
\end{abstract}



\maketitle

\section{Introduction}

Learning computer programming is a well-known challenge in undergraduate Computer Science education \cite{Watson2014, martins2018problem}. Although numerous educational technologies and simplified environments have been developed to support novice learners, the cognitive complexity of programming—particularly within the Object-Oriented Programming (OOP) paradigm—remains significant. Abstract concepts such as encapsulation, inheritance, and polymorphism often pose difficulties that go beyond syntactic understanding, affecting not only program correctness but also the quality of software design \cite{Moser_1997}.

Several studies \cite{keuningqualityeducationmapping, Gutierrez_2022, chamorro2020analysis} have shown that many students reach advanced programming courses with an insufficient conceptual grasp of OOP fundamentals, often resulting in low-quality code and persistent misconceptions. While introductory courses tend to emphasize syntactic and functional correctness, structural and stylistic aspects are frequently overlooked. As a result, students may develop ineffective mental models and poor programming practices that endure throughout their academic journey \cite{Mazaitis1993}.

A review of the literature reveals a limited number of studies that classify the specific difficulties in learning OOP. Gutiérrez, Guerrero, and López-Ospina \cite{Gutierrez_2022} identified 14 categories of such difficulties based on a review of 56 studies. Other works, such as those by Thomasson, Ratcliffe, and Thomas \cite{thomasson2006identifying}, Biju \cite{biju2013difficulties}, Holland, Griffiths, and Woodman \cite{holland97}, and Or-Bach and Lavy \cite{lavy_orbach}, organize learning challenges around fundamental OOP concepts: abstraction, encapsulation, inheritance, and polymorphism \cite{sommerville}.

Furthermore, in the context of OOP learning, it is common to evaluate students' source code based solely on whether it meets functional requirements, even if the implementation is poor and filled with \textit{code smells}. Although such code may be accepted by the compiler, it often violates key OOP principles. In many introductory courses, these design-related issues are ignored to prevent overwhelming inexperienced learners with excessive complexity. However, in OOP, it is essential not only to check if the code runs correctly, but also to assess whether it adheres to the paradigm's design principles.

Therefore, it is important not only to identify deficiencies in OOP learning but also to understand the underlying factors that contribute to them. One effective approach is source code analysis, through which signs of OOP misuse can reveal the specific learning challenges students may be facing.

This study aims to map the relationship between indicators of problematic object-oriented code and common learning difficulties in OOP. To achieve this, we conduct a qualitative analysis grounded in theories of code smells and the SOLID principles, establishing connections between these design violations and documented challenges in learning OOP \cite{Gutierrez_2022}. 

The main contributions of this work are twofold. First, we refine and extend the learning challenges previously identified by \cite{Gutierrez_2022} by defining specific categories of learning problems associated with each challenge. Second, we present a conceptual map that illustrates how particular design flaws in students' code may reflect underlying learning difficulties related to core object-oriented principles.

\section{Theoretical Background}

This section provides the theoretical background that supports this study. It addresses three key aspects: the main challenges associated with learning programming—particularly object-oriented programming, the role of code quality in programming education, and the identification of code-level indicators that may reflect learning difficulties in OOP contexts.

\subsection{Programming Learning and Challenges}
\label{sec:aprendizagem}

Moser~\cite{Moser_1997} describes programming as an intimidating process requiring multilayered skill development. Learning progresses bottom-up, starting with syntax, then structure, and finally style. Tan, Ting, and Ling~\cite{tan2009learning} note that early focus on syntax can lead to misunderstandings of deeper programming concepts, resulting in a reliance on specific languages rather than general programming skills. This often leads to poor code quality and difficulty transitioning to other languages.
Several studies have analyzed learning difficulties in programming~\cite{Moser_1997, tan2009learning, Watson2014, martins2018problem, konecki2014problems, Mazaitis1993, keuningqualityeducationmapping}. Lahtinen, Ala-Mutka, and Järvinen~\cite{lahtinen2005} collected feedback from over 500 students worldwide, confirming the widespread difficulty of programming education, particularly in abstraction and program construction.
Cheah~\cite{cheah2020factors}, through a literature review, identified key factors contributing to these challenges: lack of logical foundations, use of industry-focused tools unsuited for learning, and high levels of anxiety. A critical issue is the formation of incorrect mental models, which leads to design flaws and bugs.
Focusing more narrowly on OOP, few studies have explored specific learning challenges. Ismail, Ngah, and Umar~\cite{ismail2010instructional} argue that OOP instruction requires different strategies than procedural or structured programming, as conventional techniques like pseudocode and flowcharts are insufficient.

Kölling~\cite{kolling1999problem1} criticizes the typical pedagogical sequencing, where procedural programming precedes OOP, reinforcing the misconception that OOP is merely an additional feature. He emphasizes that OOP is a distinct paradigm that fundamentally reshapes how problems are modeled and should be taught from the outset.
Among the reviewed works, only Gutiérrez, Guerrero, and López-Ospina~\cite{Gutierrez_2022} provide a classification of OOP-specific learning difficulties. Their study defines 14 categories of student learning challenges, offering a structured framework that serves as the basis for the OOP learning challenges adopted in this research.

\subsection{Code Quality in Programming Education}

Software quality and code quality are closely related but distinct concepts. Code quality is understood as a more specific aspect of the broader software quality defined in ISO/IEC 25000~\cite{ISO25000}. The notion of code quality~\cite{keuningqualityeducationmapping} focuses on static characteristics of programs that are directly observable from the source code, excluding aspects related to dynamic behavior, such as runtime performance.
In the educational context, Stegeman, Barendsen, and Smetsers~\cite{stegeman} proposed a model with six criteria to assess student-written code. These criteria, developed from best practice guides and programming instructors' experience, include:
\begin{itemize}
  \item \textbf{Comments}: content and summarization;
  \item \textbf{Formatting}: consistency and expressiveness;
  \item \textbf{Layout}: cohesion, organization, and presence of dead code;
  \item \textbf{Naming}: consistency and meaningfulness;
  \item \textbf{Structure}: abstraction, duplication, modularization, type use, method adequacy, and fragmentation;
  \item \textbf{Expressiveness}: phrasing, clarity, and flow control.
\end{itemize}

Other studies have also contributed to evaluation criteria for code quality in education. Hamm et al.~\cite{hamm1983tool} emphasize documentation, structure, and functionality. Howatt~\cite{howatt1994criteria} proposes evaluating executability, adherence to requirements, effective comments, readability, and planning.

Based on these and other works~\cite{becker2003grading, smith2005weighted}, it is evident that code quality is a well-established topic in programming education. However, no studies were found that specifically parameterize the evaluation of code quality in the context of OOP. Although general quality criteria may apply, they often fail to address specific aspects of OOP, which introduces distinct concepts that must be validated in educational settings.
While summative evaluation through classification and conformance checking plays a role in education, such approaches offer limited support for formative learning. As this study aims to propose a formative rather than summative strategy, it focuses on identifying the weaknesses that lead to code quality problems.

\subsection{Indicators of Code Quality Issues}

Identifying issues in object-oriented code often requires the detection of recurring structural and behavioral patterns that undermine readability, maintainability, and extensibility. Key indicators of such issues include code smells \cite{fowler2018refactoring}, as well as violations of widely accepted design principles, such as the Law of Demeter \cite{demeter}, the Tell, Don't Ask principle \cite{thomas2019pragmatic}, and the SOLID principles \cite{martin2006agile}. These violations not only reflect poor code quality but also suggest fundamental misunderstandings of core object-oriented programming concepts like encapsulation, abstraction, and responsibility. Identifying these issues in students' source  code can reveal several indicators of difficulties in grasping fundamental OOP concepts.

Among these issues, code smells are perhaps the most extensively discussed in the literature, a term popularized by Fowler’s work \cite{fowler2018refactoring}. A "smell" refers to an underlying problem in the software, which can manifest at both the code level \cite{fowler2018refactoring} and the design level \cite{brown1998antipatterns}. These "smells" are symptoms in software components that can hinder the system's evolution. Depending on the level of abstraction, they are classified as code smells or design smells. Unlike bugs, which often result in immediate faults, smells do not directly cause application errors but can lead to long-term negative consequences, such as difficulties in maintenance and future development.

Several authors have contributed to the conceptualization of code smells. Brown et al. \cite{brown1998antipatterns} introduced 40 anti-patterns, which describe common problems that typically result in negative consequences. Fowler \cite{fowler2018refactoring} cataloged 22 distinct smells and proposed sequences of refactorings to mitigate each one. Wake \cite{wake2004refactoring} explored problematic patterns commonly identified by practitioners in the field. Kerievsky \cite{kerievsky2005refactoring} expanded this discussion, focusing on the role of design patterns in addressing these issues. 

An effective way to improve understanding of code smells is through their categorization based on potential relationships, which can support deeper comprehension and analysis. One of the most accepted classifications of code smells is presented by \cite{mantyla2003taxonomy}, who introduced a detailed taxonomy, grouping smells into the following categories:
\begin{itemize}
    \item Bloaters: These are code elements that have grown excessively large and become difficult to manage or understand.
    \item Object-Orientation Abusers: These represent an improper or suboptimal use of object-oriented principles, often involving workarounds that ignore good OO design practices.
    \item Change Preventers: Structures that make software modifications difficult, increasing the cost and risk of changes.
    \item Dispensables: Code fragments that are unnecessary and should be removed to improve clarity and maintainability.
    \item Couplers: Smells that indicate excessive coupling between classes or components, which can reduce modularity and hinder reuse.
\end{itemize}

Furthermore, over the years, the concept of code smells has expanded beyond traditional object-oriented code, with research identifying smells in various domains. These include test code \cite{greiler2013automated, hauptmann2013hunting} aspect-oriented systems \cite{alves2014avoiding, macia2011exploratory}, software reuse \cite{long2001software}, and web applications \cite{nguyen2012detection} among others.

Another important conceptual framework for identifying design issues in object-oriented code is \textbf{SOLID}, an acronym for five principles introduced by Martin~\cite{martin2006agile} to promote robust and maintainable software design. These principles are: (1) the \textit{Single Responsibility Principle} (SRP), which states that a class should have only one reason to change; (2) the \textit{Open/Closed Principle} (OCP), which advocates for designing modules that are open to extension but closed to modification; (3) the \textit{Liskov Substitution Principle} (LSP), which ensures that subclasses can be substituted for their base classes without compromising program correctness~\cite{liskovsubstitution}; (4) the \textit{Interface Segregation Principle} (ISP), which encourages the creation of small, role-specific interfaces rather than large, general-purpose ones; and (5) the \textit{Dependency Inversion Principle} (DIP), which promotes depending on abstractions rather than concrete implementations.

In this work, we are specifically interested in indicators related to implementation and design. For this reason, we focus on \textit{code smells} and the \textit{SOLID} principles, as summarized in Table~\ref{table:indicators}.

\begin{table}[h]
\centering
\caption{Code quality indicators classified by focus and type}
\label{table:indicators}
\begin{tabular}{lll}
\hline
\textbf{Focus} & \textbf{Type} & \textbf{References} \\
\hline
Implementation & Code Smell & \cite{fowler2018refactoring, riel1996object, brown1998antipatterns, vidal2016approach} \\
\hline
Design & SOLID & \cite{martin2006agile} \\
       & Code Smell & \cite{suryanarayana2014refactoring, marquardt2001dependency} \\
\hline
\end{tabular}
\end{table}

\section{Research Structure}\label{research}

\begin{figure*}
 \includegraphics[width=\textwidth]{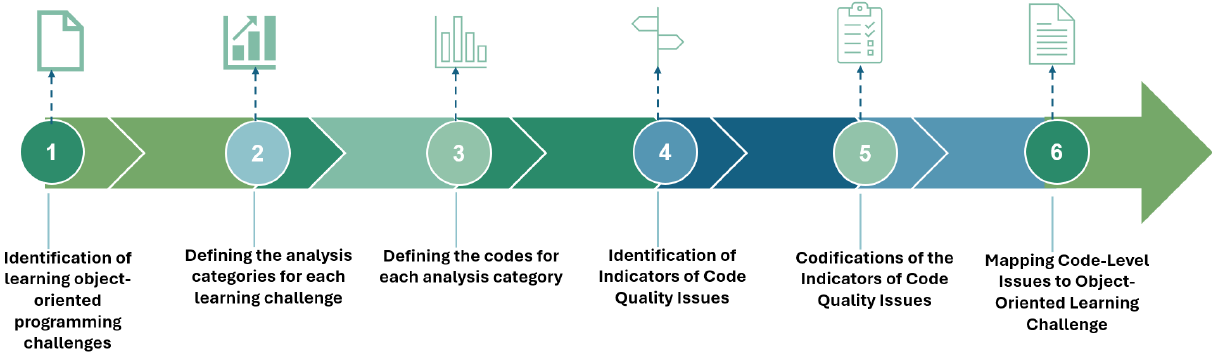}
  \caption{The Research Structure of the Study}
  \label{fig:method}
\end{figure*}

The schematic representation of the research structure applied in the study is shown in Figure \ref{fig:method}, which illustrates the stages and the output artifacts produced in each phase. The study unfolded in six key phases, starting with the identification of challenges in learning object-oriented programming and culminating in a mapping between code-level issues and object-oriented learning challenges.

\subsection{Identification of learning object-oriented programming challenges}

In the study by \cite{Gutierrez_2022}, a systematic review was conducted, in which 56 selected studies were analyzed, leading to the identification of 14 challenges related to the teaching and learning of object-oriented programming. The work of \cite{Gutierrez_2022} consolidated the main difficulties encountered in learning object-oriented programming, providing a comprehensive overview of the topic based on the existing literature. This study serves as the starting point for our mapping.

Among the 14 difficulties identified, several are not directly observable in source code or are primarily related to the teaching process rather than the learning process. Examples include the Difficulty in teaching and understanding general programming topics (D09) and the Difficulty with project administration and management methodologies and techniques (D13). Considering this in our mapping, we focused on six learning challenges that can be identified through source code, as presented by \cite{Gutierrez_2022} and supported by many works as presented below:
\begin{enumerate}
    \item \textbf{Difficulties related to understanding classes (D02)}. This difficulty is described as the complexity presented by the students when assimilating the static nature and depth of classes. It is challenging for them to understand the hierarchy and the identification of correct classes. The students even refer to the difficulty in distinguishing between class and object. They generally assimilate class as a collection of objects, rather 
    than an abstraction 
    \cite{benander2004factors, lewis2004experts, moussa2016proposing, rajashekharaiah2016design, sheetz1997exploring, tegarden2001cognitive, yang2018evaluations, gorschek2010large, hubwieser2011students, karahasanovic2007comprehension, sanders2008student, sien2012threshold}.

    \item \textbf{Difficulty in understanding the concept of method (D03)}. In this case it is referred as the complexity presented when assimilating the concept of method, there is no clarity on how to make the method calls. The students do not know how to determine the number of methods needed or what labels or names to assign to them \cite{gorschek2010large, hubwieser2011students, karahasanovic2007comprehension, moussa2016proposing, sanders2008student, sheetz1997exploring, tegarden2001cognitive}.

 \item \textbf{Difficulty in implementing object-orientation (D04)}. This problem is specified as the challenge of performing object-oriented analysis, design, and programming. The students present difficulties when adopting the object-oriented paradigm, because their initial formative process is generally based on purely structural programming. The modular nature of the object-oriented paradigm is conceived as a challenge for educators, since in this process it is common for students to assimilate erroneous conceptions and to present problems in understanding and implementing object-oriented standards \cite{benander2004factors, lewis2004experts, moussa2016proposing, rajashekharaiah2016design, sheetz1997exploring, tegarden2001cognitive, yang2018evaluations, gorschek2010large, hubwieser2011students, karahasanovic2007comprehension, sanders2008student, sien2012threshold}.
  \item \textbf{Difficulty in understanding object-oriented relationships (D05)}. It refers to the difficulty that the students have when understanding and implementing object-oriented relationships, such as association, dependency, generalization / specialization-inheritance, composition and aggregation. These problems are common due to the learners’ lack of experience in relation to the object-oriented programming paradigm. The students generally present difficulties in the process of modeling these relationships, and consequently in the implementation and application of concepts that are often conceived as complex 
  \cite{benander2004factors, gorschek2010large, hadar2013intuition, karahasanovic2007comprehension, lewis2004experts, moussa2016proposing, rajashekharaiah2016design, sheetz1997exploring, sien2012threshold, tegarden2001cognitive, yang2018evaluations}.
 \item \textbf{Difficulty in understanding polymorphism and overload (D06)}. In this case it is indicated the high level of complexity the concepts of polymorphism and overload have at the moment of initiating a student into the programming area \cite{benander2004factors, lewis2004experts, moussa2016proposing, rajashekharaiah2016design, sheetz1997exploring, tegarden2001cognitive, yang2018evaluations}. 
 \item \textbf{Difficulty in understanding encapsulation (D07)}. This problem is related to the  assimilation of several misconceptions related to understanding encapsulation, modularity and information hiding \cite{gorschek2010large, hubwieser2011students, karahasanovic2007comprehension, lewis2004experts, moussa2016proposing, rajashekharaiah2016design, sanders2008student, sheetz1997exploring, sien2012threshold, tegarden2001cognitive, xinogalos2015object, yang2018evaluations}.
 \end{enumerate}

\subsection{Defining the analysis categories for each learning challenge}

To define the analysis categories, we employed content analysis, a qualitative research method. Content analysis systematically examines the content and structure of communication, aiming to identify patterns, themes, and relationships within the data \cite{bardin2011content, krippendorff2018content}. It also allows for inferences by interpreting evidence and indicators, supported by a structured framework for technical validation \cite{bardin2011content}.

\begin{figure}
    \centering
    \includegraphics[width=0.9\linewidth]{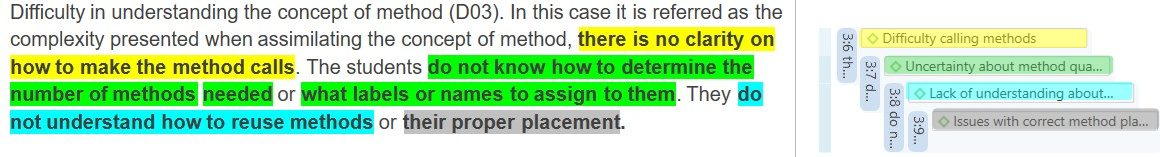}
    \caption{Identification of analysis categories for the learning challenge \textit{Difficulty in understanding the concept of method (D03)}.}
    \label{fig:categories}
\end{figure}

At this stage, we applied an inductive approach, which involves an open coding process in which categories are created during the analysis. We examined the textual descriptions of each learning challenge and defined distinct analysis categories accordingly. For example, as illustrated in Figure~\ref{fig:categories}, we identified four analysis categories for the challenge \textit{ D03 (Difficulty in understanding the concept of a method)}. These categories reflect specific issues that may contribute to the learning difficulty. In this analysis process, we used the Atlas.ti \footnote{Atlas.ti  https://atlasti.com} software to support and organize the analysis, and this process was repeated for the six learning challenges, where we have identified 22 categories.

\subsection{Defining the codes  for each analysis category} \label{lab:defining codes}

\begin{figure}
    \centering
    \includegraphics[width=0.9\linewidth]{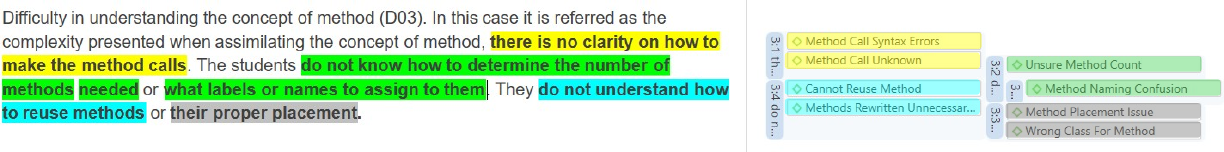}
    \caption{Identification of codes for the learning challenge \textit{Difficulty in understanding the concept of method (D03)}.}
    \label{fig:codes}
\end{figure}

\begin{figure}
    \centering
    \includegraphics[width=0.9\linewidth]{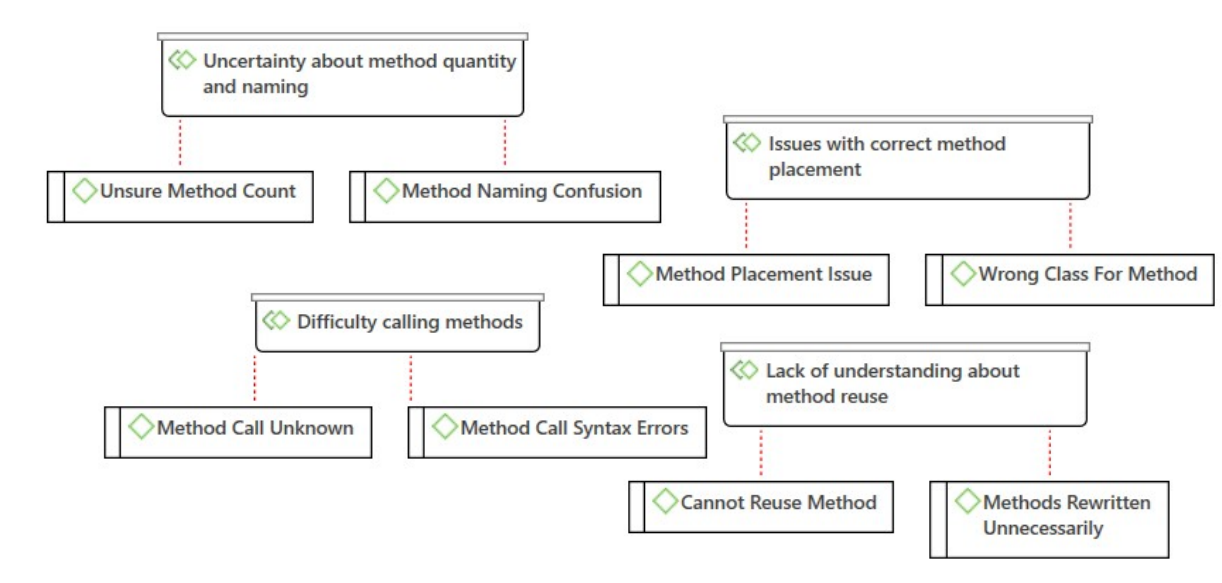}
    \caption{Analysis categories and its codes for the learning challenge \textit{Difficulty in understanding the concept of method (D03)}.}
    \label{fig:categoriesandcodes}
\end{figure}

The next stage involved defining codes for each category. Once again, we adopted an inductive approach through an open coding process. In this process, codes were created for each category. As an example, Figure~\ref{fig:codes} presents the codes identified from the text describing the learning challenge \textit{Difficulty in understanding the concept of method (D03)}, organized according to each category. As a result of the analysis of challenge D03, we identified eight codes associated with the four analysis categories defined for this challenge, as presented in Figure~\ref{fig:categoriesandcodes}. We repeated this process systematically for all other learning problems. 

\subsection{Identification of Indicators of Code Quality Issues}

The analysis of students' source code can reveal a range of indicators that point to difficulties in assimilating the fundamental concepts of OOP. Some of these indicators are reflected in the presence of code smells as well as in violations of recognized design principles such as the Law of Demeter \cite{demeter}, the Tell, Don't Ask principle \cite{thomas2019pragmatic}, and the SOLID principles  \cite{martin2006agile}. The occurrence of these violations, in addition to representing a code quality issue, indicates challenges in learning the core concepts of OOP. Such violations suggest an insufficient understanding of encapsulation, abstraction, responsibility, and other pillars of object orientation.

Considering the code problem indicators identified in the literature, many of them tend to overlap, as they address similar aspects. In this work, we selected a set of implementation and design code smells, as well as violations of SOLID principles, as indicators. Based on the literature, we analyzed the most relevant implementation and design smells and SOLID principles, and, drawing on the work of \cite{sharma2018survey, lacerda2020code} we compiled a list of code indicators that may point to difficulties in learning object-oriented programming concepts as presented in Table \ref{tab:code_smells}.

\begin{table*}[ht]
\centering
\caption{Code-Based Indicators of Difficulties in Learning Object-Oriented Programming Used. Adapted from \cite{sharma2018survey}}
\label{tab:code_smells}
\scriptsize
\begin{tabular}{|p{5cm}|p{12cm}|}
\hline
\textbf{Indicators of Object-Oriented Learning Challenges} & \textbf{Description} \\
\hline
Large class \cite{fowler2018refactoring} & A class that centralizes too many responsibilities, violating the Single Responsibility Principle. \textbf{Related} Insufficient modularization \cite{suryanarayana2014refactoring}, Blob \cite{brown1998antipatterns}, Brain class \cite{vidal2016approach} - God Class \cite{riel1996object}, Single Responsibility Principle \cite{martin2006agile}. \\
\hline
Feature envy \cite{fowler2018refactoring} & A method that accesses data from another object more than from its own class. \\
\hline
Shotgun surgery \cite{fowler2018refactoring} & A single change requires modifications in many different classes simultaneously. \\
\hline
Data class \cite{fowler2018refactoring} & A class that contains only fields and accessors with little or no behavior. \\
\hline
Long method \cite{fowler2018refactoring} & A method that is too long and complex, making it hard to understand or maintain. \textbf{Related} Broken modularization \cite{suryanarayana2014refactoring}, Single Responsibility Principle \cite{martin2006agile}. \\
\hline
Functional decomposition \cite{brown1998antipatterns} & Procedural-style code in OO programming that lacks true object orientation. \\
\hline
Refused bequest \cite{fowler2018refactoring} & A subclass inherits methods or data it doesn't need or use. \textbf{Related} Rebellious hierarchy \cite{suryanarayana2014refactoring}, Liskov Substitution Principle \cite{martin2006agile}. \\
\hline
Spaghetti code \cite{brown1998antipatterns} & Code with tangled logic and flow, making it difficult to follow and modify. \\
\hline
Divergent change \cite{fowler2018refactoring} & A class that is often changed in different ways for different reasons. \textbf{Related} Multifaceted abstraction \cite{suryanarayana2014refactoring}, Single Responsibility Principle \cite{martin2006agile}. \\
\hline
Long parameter list \cite{fowler2018refactoring} & A method that takes too many parameters, making it hard to use and refactor. \\
\hline
Duplicate code \cite{fowler2018refactoring} & Identical or very similar code exists in more than one place. \textbf{Related} Duplicate abstraction \cite{suryanarayana2014refactoring}, Unfactored hierarchy \cite{suryanarayana2014refactoring}, Cut and paste programming \cite{brown1998antipatterns}. \\
\hline
Cyclically-dependent modularization \cite{suryanarayana2014refactoring} & Modules that depend on each other in a circular way, harming modularity. \textbf{Related} Dependency cycles \cite{marquardt2001dependency}. \\
\hline
Deficient encapsulation \cite{suryanarayana2014refactoring} & Internal implementation details are exposed, reducing flexibility and safety. \\
\hline
Speculative generality \cite{fowler2018refactoring} & Code designed for future needs that may never occur, adding unnecessary complexity. \textbf{Related} Speculative hierarchy \cite{suryanarayana2014refactoring}, Open/Closed Principle \cite{martin2006agile}. \\
\hline
Lazy class \cite{fowler2018refactoring} & A class that does too little to justify its existence. \textbf{Related} Unnecessary abstraction \cite{suryanarayana2014refactoring}. \\
\hline
Switch statement \cite{fowler2018refactoring} & Complex conditional logic spread through code instead of using polymorphism. \textbf{Related} Complicated Boolean Expression \cite{wake2004refactoring}, Conditional Complexity \cite{kerievsky2005refactoring}, Unexploited encapsulation \cite{suryanarayana2014refactoring}, Missing hierarchy \cite{suryanarayana2014refactoring}, Repeated Switches \cite{fowler2018refactoring}, Open/Closed Principle \cite{martin2006agile}. \\
\hline
Primitive obsession \cite{fowler2018refactoring} & Overuse of primitive types instead of creating small objects for concepts. \textbf{Related} Missing abstraction \cite{suryanarayana2014refactoring}. \\
\hline
Swiss army knife \cite{brown1998antipatterns} & A class with too many unrelated responsibilities or utilities. \textbf{Related} Multifaceted abstraction \cite{suryanarayana2014refactoring}. \\
\hline
Data Clump \cite{fowler2018refactoring} & Groups of variables that appear together repeatedly and should be encapsulated. \\
\hline
Inappropriate Intimacy \cite{fowler2018refactoring} & Classes that know too much about each other's internals. \\
\hline
Temporary Field \cite{fowler2018refactoring} & Instance variables that are only sometimes used, depending on the context. \\
\hline
Middle Man \cite{fowler2018refactoring} & A class that delegates all work to another class, adding unnecessary indirection. \\
\hline
Message Chains \cite{fowler2018refactoring} & Chained method calls that expose navigation through multiple objects. \\
\hline
Parallel Inheritance Hierarchies \cite{fowler2018refactoring} & Adding a subclass in one hierarchy forces changes in another related hierarchy. \\
\hline
Alternative Classes with Different Interfaces \cite{fowler2018refactoring} & Classes that perform similar work but have different interfaces, complicating usage. \\
\hline
Interface Segregation Principle \cite{martin2006agile} & Interfaces should be specific and focused. A module should not be forced to depend on methods it does not use. This avoids large, general-purpose interfaces and promotes low coupling and high cohesion.\\
\hline
Dependency Inversion Principle \cite{martin2006agile} & High-level modules should depend on abstractions, not on implementations. Details should depend on abstractions, not the other way around. This principle encourages the use of interfaces and dependency injection to reduce coupling between components.\\
\hline
\end{tabular}

\end{table*}

\subsection{Codifications of the Indicators of Code Quality Issues}

The next stage involved applying the codes to each indicator identified in Table \ref{tab:code_smells}. In this step, we employ a deductive analysis approach, in which a predefined set of categories is established, and the collected data is coded according to these categories \cite{krippendorff2018content}.

We analyze each original description of the indicator indicators of the problem and apply the codes defined in Stage 3 (Section~\ref{lab:defining codes}), in order to map the relationships between the indicator of the quality of the code and the categories of analysis of learning problems and, consequently, the associated learning difficulties.

\begin{figure}
    \centering
    \includegraphics[width=0.9\linewidth]{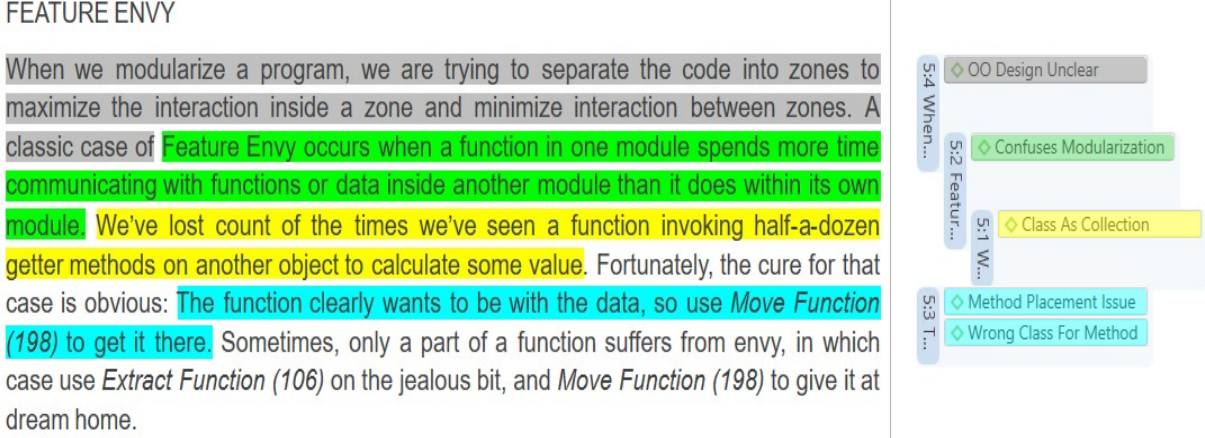}
    \caption{Codes applied to the code smell \textit{Feature Envy} \cite{fowler2018refactoring} }
    \label{fig:smell_codes}
\end{figure}

As an example, Figure~\ref{fig:smell_codes} presents the original description of the code smell \textit{Feature Envy} along with the codification applied. It shows five codes applied, corresponding to four different analysis categories. We repeat the coding process for all indicators in the Table \ref{tab:code_smells}.

\subsection{Mapping Code-Level Issues to Object-Oriented Learning Challenge}

The final stage of our research involved mapping the indicators of code quality issues to the broader challenges of learning object-oriented programming, and compiling the results presented in the following section.

\section{Results}

After executing the entire coding process, Table \ref{tab:oop_difficulties} summarizes one of the main results of the study. It organizes the identified difficulties in learning Object-Oriented Programming into analysis categories, each associated with specific codes (registration units). Additionally, for each code, related code smells and design issues are listed, reflecting how these learning difficulties can manifest in students' source code.

\begin{table*}[htbp]
\centering
\caption{Challenges in OOP learning associated with code issues}
\label{tab:oop_difficulties}
\footnotesize
\begin{tabularx}{\textwidth}{|>{\raggedright\arraybackslash}p{2.8cm}|p{3.8cm}|p{4cm}|X|}
\hline
\textbf{Learning Challenges} & \textbf{Analysis Category} & \textbf{Codes (Registration Units)} & \textbf{Code Issues} \\
\hline

\multirow{5}{*}{D02 Classes} & Static nature of class & D02.01 Static Not Assimilated & \textit{Speculative Generality}, \textit{Data Class} \\
                          & Understanding class hierarchy & D02.02 No Hierarchy, D02.03 Wrong Abstraction Level & \textit{Refused Bequest}, \textit{Large Class}, \textit{Cyclically-dependent modularization}, \textit{Parallel Inheritance Hierarchies} \\
                          & Identifying correct classes & D02.04 Cannot Identify Classes, D02.05 RealWorld Obj As Class & \textit{Large Class}, \textit{Shotgun Surgery}, \textit{Swiss army knife}, \textit{Alternative Classes with Different Interfaces}, \textit{Interface Segregation Principle} \\
                          & Confusion between class and object & D02.06 Class Equals Object & \textit{Data Class}, \textit{Large Class}, \textit{Deficient encapsulation} \\
                          & Class perceived as a collection of objects & D02.07 Class As Collection, D02.08 No Abstraction In Class & \textit{Data Class}, \textit{Feature Envy} \\
\hline
\multirow{4}{*} {D03 – Methods} & Difficulty calling methods & D03.01 Method Call Unknown, D03.02 Method Call Syntax Errors & \textit{Message Chains}, \textit{Long Parameter List}, \textit{Data Clump} \\
& Uncertainty about method quantity and naming & D03.03 Unsure Method Count, D03.04 Method Naming Confusion & \textit{Long Method}, \textit{Inappropriate Intimacy}, \textit{Data Clump} \\
& Lack of understanding about method reuse & D03.05 Cannot Reuse Method, D03.06 Methods Rewritten Unnecessarily & \textit{Duplicated Code}, \textit{Lazy Class}, \textit{Switch Statements} \\
& Issues with correct method placement & D03.07 Method Placement Issue, D03.08 Wrong Class For Method & \textit{Feature Envy}, \textit{Divergent Change} \\
\hline

\multirow{5}{*} {D04 – OO Paradigm} & Difficulty shifting from structured to OO thinking & D04.01 Structured Thinking Dominance & \textit{Large Class}, \textit{Long Method}, \textit{Functional decomposition}, \textit{Spaghetti code} \\
& Difficulty in OO analysis and design & D04.02 OO Design Unclear & \textit{Feature Envy}, \textit{Shotgun Surgery}, \textit{Cyclically-dependent modularization} \\
& Difficulty implementing OO concepts & D04.03 OO Implementation Failures, D04.04 Confused OO Syntax & \textit{Divergent Change}, \textit{Large Class}, \textit{Primitive Obsession}, \textit{Functional decomposition}, \textit{Spaghetti code}, \textit{Swiss army knife}, \textit{Dependency Inversion Principle}, \textit{Interface Segregation Principle} \\
& Misconceptions about OO paradigm & D04.05 OO Misconceptions, D04.06 Confuses Modularization & \textit{Shotgun Surgery}, \textit{Feature Envy}, \textit{Functional decomposition}, \textit{Spaghetti code}, \textit{Swiss army knife}, \textit{Dependency Inversion Principle} \\
\hline

\multirow{4}{*}{D05 – OO Relationships} 
& Trouble with association and dependency & D05.01 Association Unclear,  D05.02 Dependency Misuse & \textit{Inappropriate Intimacy, Middle Man, Dependency Inversion Principle} \\
& Difficulty with generalization/specialization (inheritance) & D05.03 Inheritance Not Applied, D05.04 No Generalization Modeling & \textit{Refused Bequest, Temporary Field, Parallel Inheritance Hierarchies} \\
& Confusion with composition and aggregation & D05.05 Aggregation Or Composition Confusion & \textit{Inappropriate Intimacy, Large Class} \\
& Difficulty modeling and implementing relationships & D05.06 Modeling Relationships Error, D05.07 Confused Concept Map & \textit{Divergent Change, Shotgun Surgery} \\
\hline
\multirow{2}{*}{\shortstack{D06 – Polymorphism\\and Overload}} 
& Polymorphism is too abstract & D06.01 Cannot Apply Polymorphism & \textit{Switch Statements, Refused Bequest} \\
& Confusion about method overloading & D06.02 Overload Not Clear, D06.03 Duplicated Methods Instead & \textit{Long Parameter List, Duplicated Code} \\
\hline

\multirow{3}{*}{D07 – Encapsulation}
& Misunderstandings about encapsulation & D07.01 Encapsulation Misunderstood, D07.02 Encapsulation As Hiding Only & \textit{Data Class, Inappropriate Intimacy, Deficient encapsulation} \\
& Lack of understanding about modularity & D07.03 Modularity Not Applied, D07.04 Mixes Concerns In Class & \textit{Large Class, Divergent Change, Alternative Classes with Different Interfaces, Interface Segregation Principle} \\
& Problems with information hiding & D07.05 Exposes Internal State, D07.06 Uses Public Attributes & \textit{Data Class, Inappropriate Intimacy, Deficient encapsulation, Data Clump} \\
\hline

\end{tabularx}

\end{table*}

Regarding all the processes carried out and the results presented in Tables \ref{tab:code_smells} and \ref{tab:oop_difficulties}, we developed a conceptual map to represent the relationships between learning challenges and code-related issues. To improve data visualization and clarity, the map was divided into parts. 

\begin{figure*}
 \includegraphics[width=\textwidth]{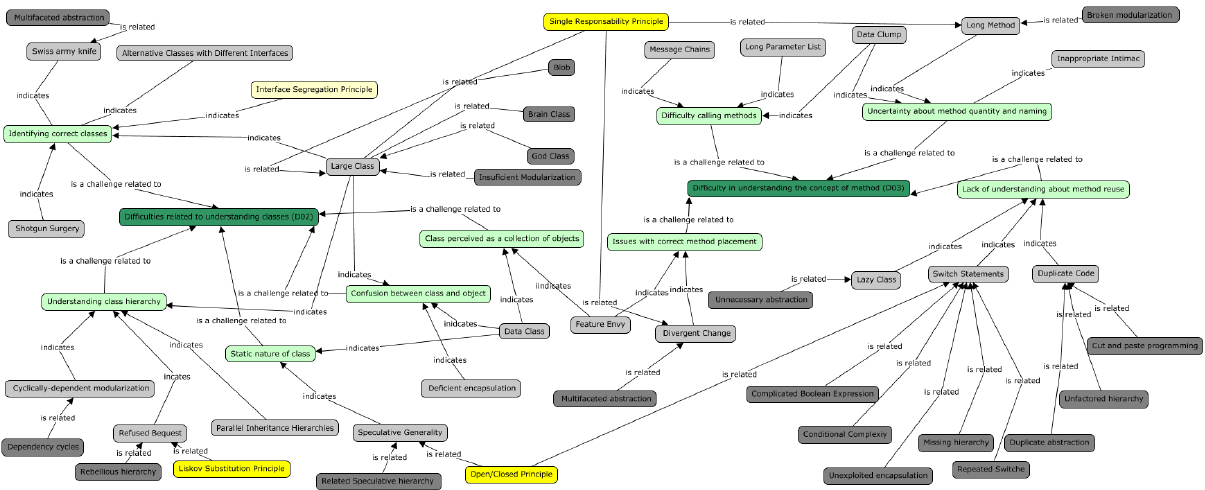}
  \caption{Conceptual map showing the relationships between learning challenges (D02 and D03) and observable code issues}
  \label{fig:map1}
\end{figure*}

\begin{figure*}
 \includegraphics[width=\textwidth]{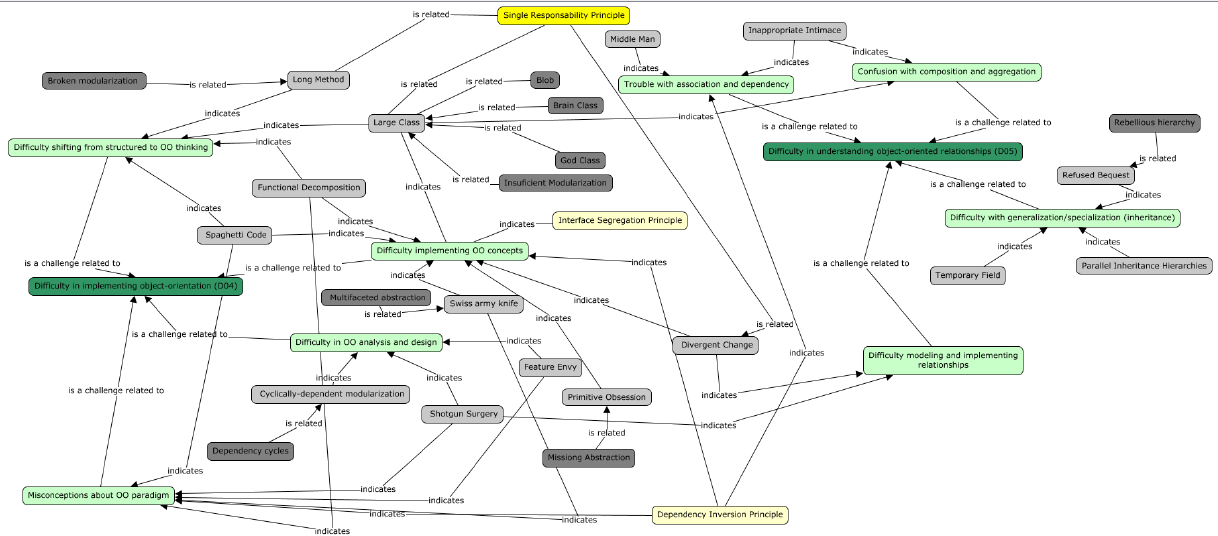}
  \caption{Conceptual map showing the relationships between learning challenges (D04 and D05) and observable code issues}
  \label{fig:map2}
\end{figure*}

\begin{figure*}
 \includegraphics[width=\textwidth]{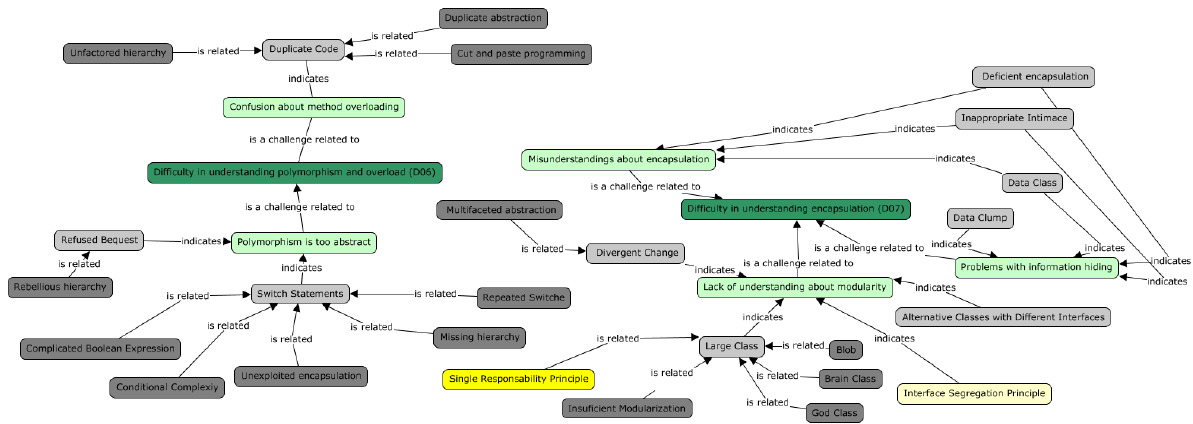}
  \caption{Conceptual map showing the relationships between learning challenges (D06 and D07) and observable code issues}
  \label{fig:map3}
\end{figure*}

Figure \ref{fig:map1} illustrates the conceptual mapping developed from the analysis process for learning challenges D02 and D03. Figure \ref{fig:map2} presents the mapping for challenges D04 and D05, while Figure \ref{fig:map3} focuses on challenges D06 and D07. The dark green elements represent the main learning challenges encountered in the context of OOP. Light green elements indicate underlying cognitive or conceptual difficulties that contribute to the emergence of these learning challenges. Observable issues in students' source code—code smells that act as indicators of such difficulties—are marked in light gray. Additionally, dark gray elements represent related code smells that, while not directly observed, are theoretically associated with the identified problems and may also signal learning difficulties. Finally, the yellow elements refer to violations of SOLID principles, which were also identified in the code and serve as indicators of design flaws linked to conceptual misunderstandings. The diagrams provides a visual representation of how abstract learning issues relate to concrete problems in students' code.

\subsection{How to Navigate in the Conceptual Map}

Through the analysis of object-oriented source code, it is possible to identify the presence of \textit{code smells} or violations of the SOLID principles. Once this analysis is conducted—either manually or with the aid of automated tools such as \textit{SonarQube}, \textit{JDeodorant}, or \textit{Checkstyle}—it becomes feasible to map students' learning difficulties.

For instance, whether the code analysis reveals the presence of the code smells \textit{Long Method} and \textit{Switch Statements}, in the Figure~\ref{fig:map1} is possible to observe that these smells are associated with the problems \textit{Uncertainty about method quantity and naming} and \textit{Lack of understanding about method reuse}, respectively. Both are indicative of the learning challenge labeled \textit{Difficulty in understanding the concept of method}. 

Additionally, \textit{Long Method} is also linked to the issue \textit{Difficulty shifting from structured to OO thinking}, which falls under the challenge \textit{Difficulty in implementing object-orientation} as presented in Figure \ref{fig:map2}. Furthermore, as shown in Figure~\ref{fig:map3}, the \textit{Switch Statement} smell is also related to the problem \textit{Polymorphism is too abstract}, which signals a \textit{Difficulty in understanding polymorphism and overload}.

The combined analysis of these two code smells suggests that the student faces broader challenges in learning object-oriented programming. These include difficulties with basic concepts such as decomposing methods into smaller units, which could also violate the Single Responsibility Principle (SRP) of SOLID, and carrying over practices from structured programming into an object-oriented context.

It is also important to highlight that the dark gray \textit{code smells} shown in Figure~\ref{fig:map1},\ref{fig:map2} and \ref{fig:map3} were not directly examined in the qualitative analysis. However, based on their known relationships with other analyzed \textit{code smells}, one can infer that their presence in a student's code may indicate associated learning difficulties. For example, if a \textit{God Class} is detected, it is plausible to infer that the student is experiencing learning difficulties similar to those linked with the \textit{Large Class} smell.

\subsection{Expert Evaluation}

To assess the educational value and practical applicability of the proposed approach, we conducted an expert evaluation involving two experienced computer science educators from Brazilian universities, with significant backgrounds in programming and software engineering instruction. One expert holds a Ph.D. and has 6 years of teaching experience, primarily in software engineering, project management, and IT governance. The other expert holds a master's degree and has 14 years of teaching experience, with expertise in software engineering, systems analysis, programming, and artificial intelligence.

Each expert was asked to review the proposed model, apply it in selected student code samples, and evaluate its ability to accurately represent real classroom challenges as well as its potential impact on teaching and learning in introductory object-oriented programming. The evaluation included a combination of structured questionnaires and open-ended feedback.

The analysis focused on the ability of the proposed mapping to faithfully reflect real learning challenges observed in the classroom. To this end, participating educators submitted authentic code excerpts from students, each containing identifiable problems commonly encountered during instruction.

Each expert then navigated the full diagnostic path proposed by our visual model: starting from the Code Issues present in the snippet, through the Analysis Categories, and finally arriving at the proposed Learning Challenges. The goal was to evaluate whether the final output corresponded meaningfully to the actual learning difficulties perceived in the classroom context for that particular student.

Table~\ref{tab:avaliacao} presents representative samples from this process, outlining the classification path and the expert’s judgment on the accuracy and relevance of the result. The samples are available at \url{https://github.com/brunostrik/BadQualityCodeExamples}

\begin{table*}[htbp]
\caption{Summary of expert evaluation}
\label{tab:avaliacao}
\centering
\footnotesize
\begin{tabularx}{\textwidth}{|c|p{2cm}|p{3cm}|p{2.3cm}|X|}
\hline
\textbf{Sample} & \textbf{Code Issues}  & \textbf{Analysis Category} & \textbf{Learning Challenges} & \textbf{Expert Judgment} \\
\hline
S1 & Divergent change, Innapropriate intimacy  & Lack of understanding about modularity, Trouble with association and dependency & D07 – Encapsulation, D05 – OO Relationships & Learning difficulties could be correctly identified through the mapping, although the problem indicator—understanding the context—was essential for identifying the correct relationships made possible by the mapping \\
\hline
S2 & Large Class, Data Class & Understanding class hierarchy, Identifying correct classes & D02 Classes & The mapping was able to identify, based on the quality issues present in the source code, a poor implementation of the class structure and the underlying learning difficulties that motivated these problems \\
\hline
S3 & Long Parameter List, Duplicated Code & Confusion about method overloading & D06 – Polymorphism and Overload & The problems caused by the incorrect implementation of method overloading were accurately mapped; however, the mapping did not capture the connection between the excessively long parameter list and the flawed class structure, which is the underlying cause of the identified issue \\
\hline
S4 & Switch Statements & Confusion about method overloading & D06 – Polymorphism and Overload & The switch statement code smell was correctly mapped as a misunderstanding of polymorphism and overloading; however, the existing codes and analysis categories could be expanded with additional elements to more clearly capture the complete absence of an appropriate polymorphic structure, as observed in the analyzed code snippet \\
\hline
\end{tabularx}

\end{table*}

The analysis carried out highlighted the strong potential of the proposed mapping to represent, in a structured way, the learning challenges faced by students in introductory object-oriented programming courses. In several cases, such as examples S1 and S2, a significant correspondence was observed between the quality issues in the students’ code and the fundamental concepts that require further development, such as encapsulation, class structuring, and understanding hierarchies. The ability to follow a clear diagnostic path—from the identified problem to the underlying conceptual difficulties—represents a meaningful step forward for pedagogical practice.

In other cases, such as examples S3 and S4, the richness of the analyzed situations helped to reveal opportunities for refinement within the model, particularly in representing certain nuances that emerge in the teaching context. Even so, the model proved to be a promising tool to support instructors in identifying and analyzing students’ difficulties. Overall, the evaluation suggests that the mapping is a useful complementary resource to the pedagogical perspective, with room to evolve as new situations and contexts are incorporated into its knowledge base.

\section{Research Limitations}
While the conceptual model presented in this study offers a novel approach to diagnosing OOP learning challenges through code analysis, it is important to acknowledge certain limitations. First, the mapping between code issues and learning difficulties was based on qualitative coding and expert interpretation.

Although we employed rigorous content analysis methods and triangulated findings with expert evaluations, subjective biases may still influence the categorization and associations identified.

Moreover, while the model incorporates widely recognized code smells and SOLID principles, it does not account for all possible design or implementation flaws that students might exhibit. Certain learning difficulties may manifest in ways not captured by the selected indicators or may result from non-code-related factors such as instructional design, student motivation, or prior knowledge.

Finally, the model’s application in classroom settings depends on educators’ familiarity with both software quality indicators and the conceptual framework. Without adequate training or supporting tools, instructors may face challenges in adopting the model effectively for formative assessment or instructional planning.

\section{Conclusion and Future Work}

Although OOP education, as well as issues such as code smells and violations of SOLID principles, are widely discussed in the literature, no studies explicitly explore the relationship between these topics. In this work, we investigated these connections through a qualitative analysis, drawing from prior studies that identify and consolidate the main learning difficulties in OOP, as well as from the literature on code smells and SOLID principles.

Our goal was to propose a conceptual map that supports the understanding of learning difficulties related to the object-oriented paradigm, based on evidence found in students' source code.  This analysis, supported by the conceptual map, enables a deeper examination of students’ difficulties in understanding core OOP concepts, rather than simply evaluating whether they are able to write functioning code. Considering this, the main contributions of the work are: 

\begin{enumerate}
    \item We identified the main object-oriented programming learning challenges that can be observed through source code analysis, based on the work of \cite{Gutierrez_2022}.
    \item We categorized learning problems associated with each of these challenges.
    \item We provided a visual representation of the relationships among different code smells identified by various authors.
    \item We identified potential learning difficulties associated with the presence of each code smell or violation of SOLID principles.
    \item We presented a visual representation that connects code smells and SOLID principle violations with object-oriented programming learning problems and their corresponding challenges.
\end{enumerate}

The expert evaluation demonstrated the model's practical applicability, with educators confirming its ability to accurately reflect real-world learning difficulties observed in student code. By linking observable code issues to underlying cognitive challenges, this work bridges the gap between software engineering practices and programming education.

To extend this research, we propose the following directions:

\begin{itemize}
    \item Tool Development: Implement automated tools to analyze student code and map issues to learning challenges, integrating with popular educational platforms.
    
    \item Expanded Language Support: Validate the model with code written in languages beyond Java, such as Python or C++, to ensure broader applicability.
    
    \item Longitudinal Studies: Investigate how early identification and intervention based on code smells impact long-term OOP proficiency.
    
    \item Instructor Dashboards: Develop dashboards to help educators track common challenges across student cohorts and tailor instruction accordingly.
\end{itemize}

This work lays the foundation for a code-centric approach to OOP education, where quality issues serve as actionable insights into student learning. Future efforts will focus on scaling its adoption in classrooms and refining the model through empirical studies.

\section*{Artifact Availability}
The samples used in the expert evaluation and the conceptual maps are available at:

https://github.com/brunostrik/BadQualityCodeExamples

\begin{acks}
Grammar and text structure were reviewed and refined with the assistance of AI tools, including Claude, Perplexity, and ChatGPT.
\end{acks}

\bibliographystyle{ACM-Reference-Format}
\bibliography{referencias.bib}


\begin{thebibliography}{61}


\ifx \showCODEN    \undefined \def \showCODEN     #1{\unskip}     \fi
\ifx \showISBNx    \undefined \def \showISBNx     #1{\unskip}     \fi
\ifx \showISBNxiii \undefined \def \showISBNxiii  #1{\unskip}     \fi
\ifx \showISSN     \undefined \def \showISSN      #1{\unskip}     \fi
\ifx \showLCCN     \undefined \def \showLCCN      #1{\unskip}     \fi
\ifx \shownote     \undefined \def \shownote      #1{#1}          \fi
\ifx \showarticletitle \undefined \def \showarticletitle #1{#1}   \fi
\ifx \showURL      \undefined \def \showURL       {\relax}        \fi
\providecommand\bibfield[2]{#2}
\providecommand\bibinfo[2]{#2}
\providecommand\natexlab[1]{#1}
\providecommand\showeprint[2][]{arXiv:#2}

\bibitem[Alves et~al\mbox{.}(2014)]%
        {alves2014avoiding}
\bibfield{author}{\bibinfo{person}{P{\'e}ricles Alves}, \bibinfo{person}{Eduardo Figueiredo}, {and} \bibinfo{person}{Fabiano Ferrari}.} \bibinfo{year}{2014}\natexlab{}.
\newblock \showarticletitle{Avoiding code pitfalls in aspect-oriented programming}. In \bibinfo{booktitle}{\emph{Programming Languages: 18th Brazilian Symposium, SBLP 2014, Maceio, Brazil, October 2-3, 2014. Proceedings 18}}. Springer, \bibinfo{pages}{31--46}.
\newblock


\bibitem[Bardin(2011)]%
        {bardin2011content}
\bibfield{author}{\bibinfo{person}{Laurence Bardin}.} \bibinfo{year}{2011}\natexlab{}.
\newblock \showarticletitle{Content analysis}.
\newblock \bibinfo{journal}{\emph{S{\~a}o Paulo: Edi{\c{c}}{\~o}es}} \bibinfo{volume}{70}, \bibinfo{number}{279} (\bibinfo{year}{2011}), \bibinfo{pages}{978--8562938047}.
\newblock


\bibitem[Becker(2003)]%
        {becker2003grading}
\bibfield{author}{\bibinfo{person}{Katrin Becker}.} \bibinfo{year}{2003}\natexlab{}.
\newblock \showarticletitle{Grading programming assignments using rubrics}. In \bibinfo{booktitle}{\emph{Proceedings of the 8th annual conference on Innovation and technology in computer science education}}. \bibinfo{pages}{253--253}.
\newblock


\bibitem[Benander et~al\mbox{.}(2004)]%
        {benander2004factors}
\bibfield{author}{\bibinfo{person}{Alan Benander}, \bibinfo{person}{Barbara Benander}, {and} \bibinfo{person}{Janche Sang}.} \bibinfo{year}{2004}\natexlab{}.
\newblock \showarticletitle{Factors related to the difficulty of learning to program in Java—an empirical study of non-novice programmers}.
\newblock \bibinfo{journal}{\emph{Information and Software Technology}} \bibinfo{volume}{46}, \bibinfo{number}{2} (\bibinfo{year}{2004}), \bibinfo{pages}{99--107}.
\newblock


\bibitem[Biju(2013)]%
        {biju2013difficulties}
\bibfield{author}{\bibinfo{person}{Soly~Mathew Biju}.} \bibinfo{year}{2013}\natexlab{}.
\newblock \showarticletitle{Difficulties in understanding object oriented programming concepts}. In \bibinfo{booktitle}{\emph{Innovations and Advances in Computer, Information, Systems Sciences, and Engineering}}. Springer, \bibinfo{pages}{319--326}.
\newblock


\bibitem[Brown et~al\mbox{.}(1998)]%
        {brown1998antipatterns}
\bibfield{author}{\bibinfo{person}{William~H Brown}, \bibinfo{person}{Raphael~C Malveau}, \bibinfo{person}{Hays W"~Skip" McCormick}, {and} \bibinfo{person}{Thomas~J Mowbray}.} \bibinfo{year}{1998}\natexlab{}.
\newblock \bibinfo{booktitle}{\emph{AntiPatterns: refactoring software, architectures, and projects in crisis}}.
\newblock \bibinfo{publisher}{John Wiley \& Sons, Inc.}
\newblock


\bibitem[Chamorro~Atalaya(2020)]%
        {chamorro2020analysis}
\bibfield{author}{\bibinfo{person}{Omar~Freddy Chamorro~Atalaya}.} \bibinfo{year}{2020}\natexlab{}.
\newblock \showarticletitle{Analysis of Learning Difficulties in Object Oriented Programming in Systems Engineering Students at UNTELS}.
\newblock  (\bibinfo{year}{2020}).
\newblock


\bibitem[Cheah(2020)]%
        {cheah2020factors}
\bibfield{author}{\bibinfo{person}{Chin~Soon Cheah}.} \bibinfo{year}{2020}\natexlab{}.
\newblock \showarticletitle{Factors contributing to the difficulties in teaching and learning of computer programming: A literature review}.
\newblock \bibinfo{journal}{\emph{Contemporary Educational Technology}} \bibinfo{volume}{12}, \bibinfo{number}{2} (\bibinfo{year}{2020}), \bibinfo{pages}{ep272}.
\newblock


\bibitem[Fowler(2018)]%
        {fowler2018refactoring}
\bibfield{author}{\bibinfo{person}{Martin Fowler}.} \bibinfo{year}{2018}\natexlab{}.
\newblock \bibinfo{booktitle}{\emph{Refactoring: improving the design of existing code}}.
\newblock \bibinfo{publisher}{Addison-Wesley Professional}.
\newblock


\bibitem[Gorschek et~al\mbox{.}(2010)]%
        {gorschek2010large}
\bibfield{author}{\bibinfo{person}{Tony Gorschek}, \bibinfo{person}{Ewan Tempero}, {and} \bibinfo{person}{Lefteris Angelis}.} \bibinfo{year}{2010}\natexlab{}.
\newblock \showarticletitle{A large-scale empirical study of practitioners' use of object-oriented concepts}. In \bibinfo{booktitle}{\emph{Proceedings of the 32nd ACM/IEEE International Conference on Software Engineering-Volume 1}}. \bibinfo{pages}{115--124}.
\newblock


\bibitem[Greiler et~al\mbox{.}(2013)]%
        {greiler2013automated}
\bibfield{author}{\bibinfo{person}{Michaela Greiler}, \bibinfo{person}{Arie Van~Deursen}, {and} \bibinfo{person}{Margaret-Anne Storey}.} \bibinfo{year}{2013}\natexlab{}.
\newblock \showarticletitle{Automated detection of test fixture strategies and smells}. In \bibinfo{booktitle}{\emph{2013 IEEE Sixth International Conference on Software Testing, Verification and Validation}}. IEEE, \bibinfo{pages}{322--331}.
\newblock


\bibitem[Gutiérrez et~al\mbox{.}(2022)]%
        {Gutierrez_2022}
\bibfield{author}{\bibinfo{person}{Luz~E. Gutiérrez}, \bibinfo{person}{Carlos~A. Guerrero}, {and} \bibinfo{person}{Héctor~A. López-Ospina}.} \bibinfo{year}{2022}\natexlab{}.
\newblock \showarticletitle{Ranking of problems and solutions in the teaching and learning of object-oriented programming}.
\newblock \bibinfo{journal}{\emph{Education and Information Technologies}} \bibinfo{volume}{27}, \bibinfo{number}{5} (\bibinfo{date}{June} \bibinfo{year}{2022}), \bibinfo{pages}{7205–7239}.
\newblock
\showISSN{1360-2357, 1573-7608}
\href{https://doi.org/10.1007/s10639-022-10929-5}{doi:\nolinkurl{10.1007/s10639-022-10929-5}}


\bibitem[Hadar(2013)]%
        {hadar2013intuition}
\bibfield{author}{\bibinfo{person}{Irit Hadar}.} \bibinfo{year}{2013}\natexlab{}.
\newblock \showarticletitle{When intuition and logic clash: The case of the object-oriented paradigm}.
\newblock \bibinfo{journal}{\emph{Science of Computer Programming}} \bibinfo{volume}{78}, \bibinfo{number}{9} (\bibinfo{year}{2013}), \bibinfo{pages}{1407--1426}.
\newblock


\bibitem[Hamm et~al\mbox{.}(1983)]%
        {hamm1983tool}
\bibfield{author}{\bibinfo{person}{R~Wayne Hamm}, \bibinfo{person}{Kenneth~D Henderson~Jr}, \bibinfo{person}{Marilyn~L Repsher}, {and} \bibinfo{person}{Kathleen~M Timmer}.} \bibinfo{year}{1983}\natexlab{}.
\newblock \showarticletitle{A tool for program grading: The Jacksonville University scale}. In \bibinfo{booktitle}{\emph{Proceedings of the fourteenth SIGCSE technical symposium on Computer science education}}. \bibinfo{pages}{248--252}.
\newblock


\bibitem[Hauptmann et~al\mbox{.}(2013)]%
        {hauptmann2013hunting}
\bibfield{author}{\bibinfo{person}{Benedikt Hauptmann}, \bibinfo{person}{Maximilian Junker}, \bibinfo{person}{Sebastian Eder}, \bibinfo{person}{Lars Heinemann}, \bibinfo{person}{Rudolf Vaas}, {and} \bibinfo{person}{Peter Braun}.} \bibinfo{year}{2013}\natexlab{}.
\newblock \showarticletitle{Hunting for smells in natural language tests}. In \bibinfo{booktitle}{\emph{2013 35th International Conference on Software Engineering (ICSE)}}. IEEE, \bibinfo{pages}{1217--1220}.
\newblock


\bibitem[Holland et~al\mbox{.}(1997)]%
        {holland97}
\bibfield{author}{\bibinfo{person}{Simon Holland}, \bibinfo{person}{Robert Griffiths}, {and} \bibinfo{person}{Mark Woodman}.} \bibinfo{year}{1997}\natexlab{}.
\newblock \showarticletitle{Avoiding object misconceptions}.
\newblock \bibinfo{journal}{\emph{ACM Sigcse Bulletin}}  \bibinfo{volume}{29}, \bibinfo{pages}{131--134}.
\newblock
\showISBNx{0897918894}
\href{https://doi.org/10.1145/268084.268132}{doi:\nolinkurl{10.1145/268084.268132}}


\bibitem[Howatt(1994)]%
        {howatt1994criteria}
\bibfield{author}{\bibinfo{person}{James~W Howatt}.} \bibinfo{year}{1994}\natexlab{}.
\newblock \showarticletitle{On criteria for grading student programs}.
\newblock \bibinfo{journal}{\emph{ACM SIGCSE Bulletin}} \bibinfo{volume}{26}, \bibinfo{number}{3} (\bibinfo{year}{1994}), \bibinfo{pages}{3--7}.
\newblock


\bibitem[Hubwieser and M{\"u}hling(2011)]%
        {hubwieser2011students}
\bibfield{author}{\bibinfo{person}{Peter Hubwieser} {and} \bibinfo{person}{Andreas M{\"u}hling}.} \bibinfo{year}{2011}\natexlab{}.
\newblock \showarticletitle{What students (should) know about object oriented programming}. In \bibinfo{booktitle}{\emph{Proceedings of the seventh international workshop on Computing education research}}. \bibinfo{pages}{77--84}.
\newblock


\bibitem[International Organization for Standardization(2014)]%
        {ISO25000}
International Organization for Standardization \bibinfo{year}{2014}\natexlab{}.
\newblock \bibinfo{booktitle}{\emph{{ISO/IEC 25000:2014 - Systems and software engineering -- Systems and software Quality Requirements and Evaluation (SQuaRE) -- Guide to SQuaRE}}}.
\newblock International Organization for Standardization, Geneva, Switzerland.
\newblock
\newblock
\shownote{Disponível em: \url{https://www.iso.org/standard/64764.html}, Acesso em: 12 ago. 2024}.


\bibitem[Ismail et~al\mbox{.}(2010)]%
        {ismail2010instructional}
\bibfield{author}{\bibinfo{person}{Mohd~Nasir Ismail}, \bibinfo{person}{Nor~Azilah Ngah}, {and} \bibinfo{person}{Irfan~Naufal Umar}.} \bibinfo{year}{2010}\natexlab{}.
\newblock \showarticletitle{Instructional strategy in the teaching of computer programming: a need assessment analyses}.
\newblock \bibinfo{journal}{\emph{The Turkish Online Journal of Educational Technology}} \bibinfo{volume}{9}, \bibinfo{number}{2} (\bibinfo{year}{2010}), \bibinfo{pages}{125--131}.
\newblock


\bibitem[Karahasanovi{\'c} et~al\mbox{.}(2007)]%
        {karahasanovic2007comprehension}
\bibfield{author}{\bibinfo{person}{Amela Karahasanovi{\'c}}, \bibinfo{person}{Annette~Kristin Levine}, {and} \bibinfo{person}{Richard Thomas}.} \bibinfo{year}{2007}\natexlab{}.
\newblock \showarticletitle{Comprehension strategies and difficulties in maintaining object-oriented systems: An explorative study}.
\newblock \bibinfo{journal}{\emph{Journal of Systems and Software}} \bibinfo{volume}{80}, \bibinfo{number}{9} (\bibinfo{year}{2007}), \bibinfo{pages}{1541--1559}.
\newblock


\bibitem[Kerievsky(2005)]%
        {kerievsky2005refactoring}
\bibfield{author}{\bibinfo{person}{Joshua Kerievsky}.} \bibinfo{year}{2005}\natexlab{}.
\newblock \bibinfo{booktitle}{\emph{Refactoring to patterns}}.
\newblock \bibinfo{publisher}{Pearson Deutschland GmbH}.
\newblock


\bibitem[Keuning et~al\mbox{.}(2023)]%
        {keuningqualityeducationmapping}
\bibfield{author}{\bibinfo{person}{Hieke Keuning}, \bibinfo{person}{Johan Jeuring}, {and} \bibinfo{person}{Bastiaan Heeren}.} \bibinfo{year}{2023}\natexlab{}.
\newblock \showarticletitle{A Systematic Mapping Study of Code Quality in Education}. In \bibinfo{booktitle}{\emph{Proceedings of the 2023 Conference on Innovation and Technology in Computer Science Education V. 1}} (Turku, Finland) \emph{(\bibinfo{series}{ITiCSE 2023})}. \bibinfo{publisher}{Association for Computing Machinery}, \bibinfo{address}{New York, NY, USA}, \bibinfo{pages}{5–11}.
\newblock
\showISBNx{9798400701382}
\href{https://doi.org/10.1145/3587102.3588777}{doi:\nolinkurl{10.1145/3587102.3588777}}


\bibitem[K{\"o}lling(1999)]%
        {kolling1999problem1}
\bibfield{author}{\bibinfo{person}{Michael K{\"o}lling}.} \bibinfo{year}{1999}\natexlab{}.
\newblock \showarticletitle{The problem of teaching object-oriented programming, Part 1: Languages}.
\newblock \bibinfo{journal}{\emph{Journal of Object-oriented programming}} \bibinfo{volume}{11}, \bibinfo{number}{8} (\bibinfo{year}{1999}), \bibinfo{pages}{8--15}.
\newblock


\bibitem[Konecki(2014)]%
        {konecki2014problems}
\bibfield{author}{\bibinfo{person}{Mario Konecki}.} \bibinfo{year}{2014}\natexlab{}.
\newblock \showarticletitle{Problems in programming education and means of their improvement}.
\newblock \bibinfo{journal}{\emph{DAAAM international scientific book}}  \bibinfo{volume}{2014} (\bibinfo{year}{2014}), \bibinfo{pages}{459--470}.
\newblock


\bibitem[Krippendorff(2018)]%
        {krippendorff2018content}
\bibfield{author}{\bibinfo{person}{Klaus Krippendorff}.} \bibinfo{year}{2018}\natexlab{}.
\newblock \bibinfo{booktitle}{\emph{Content analysis: An introduction to its methodology}}.
\newblock \bibinfo{publisher}{Sage publications}, \bibinfo{address}{Thousand Oaks, California}.
\newblock


\bibitem[Lacerda et~al\mbox{.}(2020)]%
        {lacerda2020code}
\bibfield{author}{\bibinfo{person}{Guilherme Lacerda}, \bibinfo{person}{Fabio Petrillo}, \bibinfo{person}{Marcelo Pimenta}, {and} \bibinfo{person}{Yann~Ga{\"e}l Gu{\'e}h{\'e}neuc}.} \bibinfo{year}{2020}\natexlab{}.
\newblock \showarticletitle{Code smells and refactoring: A tertiary systematic review of challenges and observations}.
\newblock \bibinfo{journal}{\emph{Journal of Systems and Software}}  \bibinfo{volume}{167} (\bibinfo{year}{2020}), \bibinfo{pages}{110610}.
\newblock


\bibitem[Lahtinen et~al\mbox{.}(2005)]%
        {lahtinen2005}
\bibfield{author}{\bibinfo{person}{Essi Lahtinen}, \bibinfo{person}{Kirsti Ala-Mutka}, {and} \bibinfo{person}{Hannu-Matti J\"{a}rvinen}.} \bibinfo{year}{2005}\natexlab{}.
\newblock \showarticletitle{A study of the difficulties of novice programmers}. In \bibinfo{booktitle}{\emph{Proceedings of the 10th Annual SIGCSE Conference on Innovation and Technology in Computer Science Education}} (Caparica, Portugal) \emph{(\bibinfo{series}{ITiCSE '05})}. \bibinfo{publisher}{Association for Computing Machinery}, \bibinfo{address}{New York, NY, USA}, \bibinfo{pages}{14–18}.
\newblock
\showISBNx{1595930248}
\href{https://doi.org/10.1145/1067445.1067453}{doi:\nolinkurl{10.1145/1067445.1067453}}


\bibitem[Lewis et~al\mbox{.}(2004)]%
        {lewis2004experts}
\bibfield{author}{\bibinfo{person}{Tracy~L Lewis}, \bibinfo{person}{Mary~Beth Rosson}, {and} \bibinfo{person}{Manuel~A P{\'e}rez-Qui{\~n}ones}.} \bibinfo{year}{2004}\natexlab{}.
\newblock \showarticletitle{What Do The Experts Say? teaching introductory design from an expert's perspective}.
\newblock \bibinfo{journal}{\emph{ACM SIGCSE Bulletin}} \bibinfo{volume}{36}, \bibinfo{number}{1} (\bibinfo{year}{2004}), \bibinfo{pages}{296--300}.
\newblock


\bibitem[Lieberherr and Holland(1989)]%
        {demeter}
\bibfield{author}{\bibinfo{person}{Karl~J. Lieberherr} {and} \bibinfo{person}{Ian~M. Holland}.} \bibinfo{year}{1989}\natexlab{}.
\newblock \showarticletitle{Assuring good style for object-oriented programs}.
\newblock \bibinfo{journal}{\emph{IEEE software}} \bibinfo{volume}{6}, \bibinfo{number}{5} (\bibinfo{year}{1989}), \bibinfo{pages}{38--48}.
\newblock


\bibitem[Liskov and Wing(1994)]%
        {liskovsubstitution}
\bibfield{author}{\bibinfo{person}{Barbara~H Liskov} {and} \bibinfo{person}{Jeannette~M Wing}.} \bibinfo{year}{1994}\natexlab{}.
\newblock \showarticletitle{A behavioral notion of subtyping}.
\newblock \bibinfo{journal}{\emph{ACM Transactions on Programming Languages and Systems (TOPLAS)}} \bibinfo{volume}{16}, \bibinfo{number}{6} (\bibinfo{year}{1994}), \bibinfo{pages}{1811--1841}.
\newblock


\bibitem[Long(2001)]%
        {long2001software}
\bibfield{author}{\bibinfo{person}{John Long}.} \bibinfo{year}{2001}\natexlab{}.
\newblock \showarticletitle{Software reuse antipatterns}.
\newblock \bibinfo{journal}{\emph{ACM SIGSOFT Software Engineering Notes}} \bibinfo{volume}{26}, \bibinfo{number}{4} (\bibinfo{year}{2001}), \bibinfo{pages}{68--76}.
\newblock


\bibitem[Macia~Bertran et~al\mbox{.}(2011)]%
        {macia2011exploratory}
\bibfield{author}{\bibinfo{person}{Isela Macia~Bertran}, \bibinfo{person}{Alessandro Garcia}, {and} \bibinfo{person}{Arndt Von~Staa}.} \bibinfo{year}{2011}\natexlab{}.
\newblock \showarticletitle{An exploratory study of code smells in evolving aspect-oriented systems}. In \bibinfo{booktitle}{\emph{Proceedings of the tenth international conference on Aspect-oriented software development}}. \bibinfo{pages}{203--214}.
\newblock


\bibitem[Mantyla et~al\mbox{.}(2003)]%
        {mantyla2003taxonomy}
\bibfield{author}{\bibinfo{person}{Mika Mantyla}, \bibinfo{person}{Jari Vanhanen}, {and} \bibinfo{person}{Casper Lassenius}.} \bibinfo{year}{2003}\natexlab{}.
\newblock \showarticletitle{A taxonomy and an initial empirical study of bad smells in code}. In \bibinfo{booktitle}{\emph{International Conference on Software Maintenance, 2003. ICSM 2003. Proceedings.}} IEEE, \bibinfo{pages}{381--384}.
\newblock


\bibitem[Marquardt(2001)]%
        {marquardt2001dependency}
\bibfield{author}{\bibinfo{person}{Klaus Marquardt}.} \bibinfo{year}{2001}\natexlab{}.
\newblock \showarticletitle{Dependency Structures {\~N} Architectural Diagnoses and Therapies.}. In \bibinfo{booktitle}{\emph{EuroPLoP}}. Citeseer, \bibinfo{pages}{11--52}.
\newblock


\bibitem[Martin and Martin(2006)]%
        {martin2006agile}
\bibfield{author}{\bibinfo{person}{Micah Martin} {and} \bibinfo{person}{Robert~C Martin}.} \bibinfo{year}{2006}\natexlab{}.
\newblock \bibinfo{booktitle}{\emph{Agile principles, patterns, and practices in C}}.
\newblock \bibinfo{publisher}{Pearson Education}.
\newblock


\bibitem[Martins et~al\mbox{.}(2018)]%
        {martins2018problem}
\bibfield{author}{\bibinfo{person}{Val{\'e}ria~F Martins}, \bibinfo{person}{Ilana de Almeida Souza~Concilio}, {and} \bibinfo{person}{Marcelo de Paiva~Guimar{\~a}es}.} \bibinfo{year}{2018}\natexlab{}.
\newblock \showarticletitle{Problem based learning associated to the development of games for programming teaching}.
\newblock \bibinfo{journal}{\emph{Computer Applications in Engineering Education}} \bibinfo{volume}{26}, \bibinfo{number}{5} (\bibinfo{year}{2018}), \bibinfo{pages}{1577--1589}.
\newblock


\bibitem[Mazaitis(1993)]%
        {Mazaitis1993}
\bibfield{author}{\bibinfo{person}{D. Mazaitis}.} \bibinfo{year}{1993}\natexlab{}.
\newblock \showarticletitle{The object-oriented paradigm in the undergraduate curriculum: a survey of implementations and issues}.
\newblock \bibinfo{journal}{\emph{ACM SIGCSE Bulletin}} \bibinfo{volume}{25}, \bibinfo{number}{3} (\bibinfo{year}{1993}), \bibinfo{pages}{58--64}.
\newblock
\href{https://doi.org/10.1145/165408.165432}{doi:\nolinkurl{10.1145/165408.165432}}


\bibitem[Moser(1997)]%
        {Moser_1997}
\bibfield{author}{\bibinfo{person}{Robert Moser}.} \bibinfo{year}{1997}\natexlab{}.
\newblock \showarticletitle{A fantasy adventure game as a learning environment: why learning to program is so difficult and what can be done about it}. In \bibinfo{booktitle}{\emph{Proceedings of the 2nd conference on Integrating technology into computer science education}} \emph{(\bibinfo{series}{ITiCSE ’97})}. \bibinfo{publisher}{Association for Computing Machinery}, \bibinfo{address}{New York, NY, USA}, \bibinfo{pages}{114–116}.
\newblock
\showISBNx{9780897919234}
\href{https://doi.org/10.1145/268819.268853}{doi:\nolinkurl{10.1145/268819.268853}}


\bibitem[Moussa et~al\mbox{.}(2016)]%
        {moussa2016proposing}
\bibfield{author}{\bibinfo{person}{Wejdan~Eissa Moussa}, \bibinfo{person}{Raniyah~Mutlaq Almalki}, \bibinfo{person}{Maryam~Abdulrahman Alamoudi}, {and} \bibinfo{person}{Arwa Allinjawi}.} \bibinfo{year}{2016}\natexlab{}.
\newblock \showarticletitle{Proposing a 3d interactive visualization tool for learning oop concepts}. In \bibinfo{booktitle}{\emph{2016 13th Learning and Technology Conference (L\&T)}}. IEEE, \bibinfo{pages}{1--7}.
\newblock


\bibitem[Nguyen et~al\mbox{.}(2012)]%
        {nguyen2012detection}
\bibfield{author}{\bibinfo{person}{Hung~Viet Nguyen}, \bibinfo{person}{Hoan~Anh Nguyen}, \bibinfo{person}{Tung~Thanh Nguyen}, \bibinfo{person}{Anh~Tuan Nguyen}, {and} \bibinfo{person}{Tien~N Nguyen}.} \bibinfo{year}{2012}\natexlab{}.
\newblock \showarticletitle{Detection of embedded code smells in dynamic web applications}. In \bibinfo{booktitle}{\emph{Proceedings of the 27th IEEE/ACM International Conference on Automated Software Engineering}}. \bibinfo{pages}{282--285}.
\newblock


\bibitem[Or-Bach and Lavy(2004)]%
        {lavy_orbach}
\bibfield{author}{\bibinfo{person}{Rachel Or-Bach} {and} \bibinfo{person}{Ilana Lavy}.} \bibinfo{year}{2004}\natexlab{}.
\newblock \showarticletitle{Cognitive activities of abstraction in object orientation: an empirical study}.
\newblock \bibinfo{journal}{\emph{SIGCSE Bull.}} \bibinfo{volume}{36}, \bibinfo{number}{2} (\bibinfo{date}{jun} \bibinfo{year}{2004}), \bibinfo{pages}{82–86}.
\newblock
\showISSN{0097-8418}
\href{https://doi.org/10.1145/1024338.1024378}{doi:\nolinkurl{10.1145/1024338.1024378}}


\bibitem[Rajashekharaiah et~al\mbox{.}(2016)]%
        {rajashekharaiah2016design}
\bibfield{author}{\bibinfo{person}{KMM Rajashekharaiah}, \bibinfo{person}{Manjula Pawar}, \bibinfo{person}{Mahesh~S Patil}, \bibinfo{person}{Nagaratna Kulenavar}, {and} \bibinfo{person}{GH Joshi}.} \bibinfo{year}{2016}\natexlab{}.
\newblock \showarticletitle{Design thinking framework to enhance object oriented design and problem analysis skill in Java programming laboratory: An experience}. In \bibinfo{booktitle}{\emph{2016 IEEE 4th International Conference on MOOCs, Innovation and Technology in Education (MITE)}}. IEEE, \bibinfo{pages}{200--205}.
\newblock


\bibitem[Riel(1996)]%
        {riel1996object}
\bibfield{author}{\bibinfo{person}{Arthur~J Riel}.} \bibinfo{year}{1996}\natexlab{}.
\newblock \bibinfo{booktitle}{\emph{Object-oriented design heuristics}}.
\newblock \bibinfo{publisher}{Addison-Wesley Longman Publishing Co., Inc.}
\newblock


\bibitem[Sanders et~al\mbox{.}(2008)]%
        {sanders2008student}
\bibfield{author}{\bibinfo{person}{Kate Sanders}, \bibinfo{person}{Jonas Boustedt}, \bibinfo{person}{Anna Eckerdal}, \bibinfo{person}{Robert McCartney}, \bibinfo{person}{Jan~Erik Mostr{\"o}m}, \bibinfo{person}{Lynda Thomas}, {and} \bibinfo{person}{Carol Zander}.} \bibinfo{year}{2008}\natexlab{}.
\newblock \showarticletitle{Student understanding of object-oriented programming as expressed in concept maps}. In \bibinfo{booktitle}{\emph{Proceedings of the 39th SIGCSE technical symposium on Computer science education}}. \bibinfo{pages}{332--336}.
\newblock


\bibitem[Sharma and Spinellis(2018)]%
        {sharma2018survey}
\bibfield{author}{\bibinfo{person}{Tushar Sharma} {and} \bibinfo{person}{Diomidis Spinellis}.} \bibinfo{year}{2018}\natexlab{}.
\newblock \showarticletitle{A survey on software smells}.
\newblock \bibinfo{journal}{\emph{Journal of Systems and Software}}  \bibinfo{volume}{138} (\bibinfo{year}{2018}), \bibinfo{pages}{158--173}.
\newblock


\bibitem[Sheetz et~al\mbox{.}(1997)]%
        {sheetz1997exploring}
\bibfield{author}{\bibinfo{person}{Steven~D Sheetz}, \bibinfo{person}{Gretchen Irwin}, \bibinfo{person}{David~P Tegarden}, \bibinfo{person}{H~James Nelson}, {and} \bibinfo{person}{David~E Monarchi}.} \bibinfo{year}{1997}\natexlab{}.
\newblock \showarticletitle{Exploring the difficulties of learning object-oriented techniques}.
\newblock \bibinfo{journal}{\emph{Journal of Management Information Systems}} \bibinfo{volume}{14}, \bibinfo{number}{2} (\bibinfo{year}{1997}), \bibinfo{pages}{103--131}.
\newblock


\bibitem[Sien and Chong(2012)]%
        {sien2012threshold}
\bibfield{author}{\bibinfo{person}{Ven Sien} {and} \bibinfo{person}{David Chong}.} \bibinfo{year}{2012}\natexlab{}.
\newblock \showarticletitle{Threshold concepts in object-oriented modelling}.
\newblock \bibinfo{journal}{\emph{Electronic Communications of the EASST}}  \bibinfo{volume}{52} (\bibinfo{year}{2012}).
\newblock


\bibitem[Smith and Cordova(2005)]%
        {smith2005weighted}
\bibfield{author}{\bibinfo{person}{Lon Smith} {and} \bibinfo{person}{Jose Cordova}.} \bibinfo{year}{2005}\natexlab{}.
\newblock \showarticletitle{Weighted primary trait analysis for computer program evaluation}.
\newblock \bibinfo{journal}{\emph{Journal of Computing Sciences in Colleges}} \bibinfo{volume}{20}, \bibinfo{number}{6} (\bibinfo{year}{2005}), \bibinfo{pages}{14--19}.
\newblock


\bibitem[Sommerville(2015)]%
        {sommerville}
\bibfield{author}{\bibinfo{person}{Ian Sommerville}.} \bibinfo{year}{2015}\natexlab{}.
\newblock \bibinfo{booktitle}{\emph{Software Engineering} (\bibinfo{edition}{10th} ed.)}.
\newblock \bibinfo{publisher}{Pearson}.
\newblock
\showISBNx{0133943038}


\bibitem[Stegeman et~al\mbox{.}(2014)]%
        {stegeman}
\bibfield{author}{\bibinfo{person}{Martijn Stegeman}, \bibinfo{person}{Erik Barendsen}, {and} \bibinfo{person}{Sjaak Smetsers}.} \bibinfo{year}{2014}\natexlab{}.
\newblock \showarticletitle{Towards an empirically validated model for assessment of code quality}. In \bibinfo{booktitle}{\emph{Proceedings of the 14th Koli Calling international conference on computing education research}}. \bibinfo{pages}{99--108}.
\newblock


\bibitem[Suryanarayana et~al\mbox{.}(2014)]%
        {suryanarayana2014refactoring}
\bibfield{author}{\bibinfo{person}{Girish Suryanarayana}, \bibinfo{person}{Ganesh Samarthyam}, {and} \bibinfo{person}{Tushar Sharma}.} \bibinfo{year}{2014}\natexlab{}.
\newblock \bibinfo{booktitle}{\emph{Refactoring for software design smells: managing technical debt}}.
\newblock \bibinfo{publisher}{Morgan Kaufmann}.
\newblock


\bibitem[Tan et~al\mbox{.}(2009)]%
        {tan2009learning}
\bibfield{author}{\bibinfo{person}{Phit-Huan Tan}, \bibinfo{person}{Choo-Yee Ting}, {and} \bibinfo{person}{Siew-Woei Ling}.} \bibinfo{year}{2009}\natexlab{}.
\newblock \showarticletitle{Learning difficulties in programming courses: undergraduates' perspective and perception}. In \bibinfo{booktitle}{\emph{2009 International Conference on Computer Technology and Development}}, Vol.~\bibinfo{volume}{1}. IEEE, \bibinfo{pages}{42--46}.
\newblock


\bibitem[Tegarden and Sheetz(2001)]%
        {tegarden2001cognitive}
\bibfield{author}{\bibinfo{person}{David~P Tegarden} {and} \bibinfo{person}{Steven~D Sheetz}.} \bibinfo{year}{2001}\natexlab{}.
\newblock \showarticletitle{Cognitive activities in OO development}.
\newblock \bibinfo{journal}{\emph{International Journal of Human-Computer Studies}} \bibinfo{volume}{54}, \bibinfo{number}{6} (\bibinfo{year}{2001}), \bibinfo{pages}{779--798}.
\newblock


\bibitem[Thomas and Hunt(2019)]%
        {thomas2019pragmatic}
\bibfield{author}{\bibinfo{person}{David Thomas} {and} \bibinfo{person}{Andrew Hunt}.} \bibinfo{year}{2019}\natexlab{}.
\newblock \bibinfo{booktitle}{\emph{The Pragmatic Programmer: your journey to mastery}}.
\newblock \bibinfo{publisher}{Addison-Wesley Professional}.
\newblock


\bibitem[Thomasson et~al\mbox{.}(2006)]%
        {thomasson2006identifying}
\bibfield{author}{\bibinfo{person}{Benjy Thomasson}, \bibinfo{person}{Mark Ratcliffe}, {and} \bibinfo{person}{Lynda Thomas}.} \bibinfo{year}{2006}\natexlab{}.
\newblock \showarticletitle{Identifying novice difficulties in object oriented design}.
\newblock \bibinfo{journal}{\emph{ACM SIGCSE Bulletin}} \bibinfo{volume}{38}, \bibinfo{number}{3} (\bibinfo{year}{2006}), \bibinfo{pages}{28--32}.
\newblock


\bibitem[Vidal et~al\mbox{.}(2016)]%
        {vidal2016approach}
\bibfield{author}{\bibinfo{person}{Santiago~A Vidal}, \bibinfo{person}{Claudia Marcos}, {and} \bibinfo{person}{J~Andr{\'e}s D{\'\i}az-Pace}.} \bibinfo{year}{2016}\natexlab{}.
\newblock \showarticletitle{An approach to prioritize code smells for refactoring}.
\newblock \bibinfo{journal}{\emph{Automated Software Engineering}}  \bibinfo{volume}{23} (\bibinfo{year}{2016}), \bibinfo{pages}{501--532}.
\newblock


\bibitem[Wake(2004)]%
        {wake2004refactoring}
\bibfield{author}{\bibinfo{person}{William~C Wake}.} \bibinfo{year}{2004}\natexlab{}.
\newblock \bibinfo{booktitle}{\emph{Refactoring workbook}}.
\newblock \bibinfo{publisher}{Addison-Wesley Professional}.
\newblock


\bibitem[Watson and Li(2014)]%
        {Watson2014}
\bibfield{author}{\bibinfo{person}{C. Watson} {and} \bibinfo{person}{F.~W.~B. Li}.} \bibinfo{year}{2014}\natexlab{}.
\newblock \showarticletitle{Failure rates in introductory programming revisited}. In \bibinfo{booktitle}{\emph{Proceedings of the 2014 conference on Innovation \& technology in computer science education - ITiCSE '14}}. \bibinfo{publisher}{ACM Press}, \bibinfo{address}{New York, New York, USA}, \bibinfo{pages}{39--44}.
\newblock
\href{https://doi.org/10.1145/2591708.2591749}{doi:\nolinkurl{10.1145/2591708.2591749}}


\bibitem[Xinogalos(2015)]%
        {xinogalos2015object}
\bibfield{author}{\bibinfo{person}{Stelios Xinogalos}.} \bibinfo{year}{2015}\natexlab{}.
\newblock \showarticletitle{Object-oriented design and programming: an investigation of novices’ conceptions on objects and classes}.
\newblock \bibinfo{journal}{\emph{ACM Transactions on Computing Education (TOCE)}} \bibinfo{volume}{15}, \bibinfo{number}{3} (\bibinfo{year}{2015}), \bibinfo{pages}{1--21}.
\newblock


\bibitem[Yang et~al\mbox{.}(2018)]%
        {yang2018evaluations}
\bibfield{author}{\bibinfo{person}{Jeong Yang}, \bibinfo{person}{Young Lee}, {and} \bibinfo{person}{Kai~H Chang}.} \bibinfo{year}{2018}\natexlab{}.
\newblock \showarticletitle{Evaluations of JaguarCode: A web-based object-oriented programming environment with static and dynamic visualization}.
\newblock \bibinfo{journal}{\emph{Journal of Systems and Software}}  \bibinfo{volume}{145} (\bibinfo{year}{2018}), \bibinfo{pages}{147--163}.
\newblock


\end{thebibliography}

\end{document}